\pdfoutput = 1
\documentclass[
	aps,
	prl,
	reprint,
	nofootinbib,
	nobibnotes,
	superscriptaddress,
	preprintnumbers,
	groupedaddress
]{revtex4-1}

\usepackage[utf8]{inputenc}
\usepackage[T1]{fontenc}
\usepackage{lmodern}
\usepackage{textcomp}
\usepackage{microtype}

\usepackage[abbreviations, british]{foreign}

\usepackage{mathtools}
\usepackage{mathrsfs}
\usepackage{fixmath}
\usepackage{alphabeta}
\usepackage{bm}
\usepackage{bbm}
\usepackage{physics}
\usepackage{units}
\usepackage{slashed}
\usepackage{isotope}

\usepackage{ragged2e}
\usepackage[inline]{enumitem}
\setlist{noitemsep}
\newlist{inlinelist}{enumerate*}{1}
\setlist*[inlinelist,1]{itemjoin={,\ }, itemjoin*={, and\ }, after=.}

\usepackage[dvipsnames]{xcolor}

\usepackage[subrefformat = parens]{subcaption}

\usepackage{threeparttable}
\DeclareCaptionFormat{revtex}{#1#2\justifying{#3}}
\usepackage{booktabs}
\usepackage{dcolumn}
\usepackage{graphicx}
\graphicspath{{./figures/}}
\providecommand{\tikzsetnextfilename}[1]{}

\usepackage[hidelinks, pdfencoding = auto]{hyperref}
\usepackage[noabbrev, capitalize]{cleveref}
\crefname{enumi}{point}{points}
\makeatletter\DeclareRobustCommand{\labelcrefrange}[2]{\@crefrangenostar{labelcref}{#1}{#2}}\makeatother

\let\oldcite\cite\renewcommand\cite{\unskip~\oldcite}
\let\oldeqref\eqref\renewcommand\eqref{\unskip~\oldeqref}
\let\oldparagraph\paragraph\renewcommand{\paragraph}[1]{\oldparagraph{#1:}}

\begin{document}

\title{Mapping the viable parameter space for testable leptogenesis}

\author{Marco Drewes$^1$}
\email{marco.drewes@uclouvain.be}
\author{Yannis Georis$^1$}
\email{yannis.georis@student.uclouvain.be}
\author{Juraj Klaric$^2$}
\email{juraj.klaric@epfl.ch}

\affiliation{$^1$Centre for Cosmology, Particle Physics and Phenomenology, Université catholique de Louvain, Louvain-la-Neuve B-1348, Belgium}

\affiliation{$^2$Institute of Physics, Laboratory for Particle Physics and Cosmology,
	École polytechnique fédérale de Lausanne,
	CH-1015 Lausanne,
Switzerland}

\begin{abstract}
	We for the first time map the range of active-sterile neutrino mixing angles in which leptogenesis is possible in the type I seesaw model with three heavy neutrinos with Majorana masses between 50 MeV and 70 TeV, covering the entire experimentally accessible mass range. Our study includes both, the asymmetry generation during freeze-in (ARS mechanism) and freeze-out (resonant leptotenesis) of the heavy neutrinos.
	The range of mixings for which leptogenesis is feasible is considerably larger than in the minimal model with only two heavy neutrinos and extends all the way up to the current experimental bounds. For such large mixing angles the HL-LHC could potentially observe a number of events that is large enough to compare different decay channels, a first step towards testing the hypothesis that these particles may be responsible for the origin of matter and neutrino masses.
\end{abstract}

\maketitle

\paragraph{\bf Introduction}
Right-handed neutrinos $\nu_R$ appear in many extensions of the Standard Model (SM) of particle physics
\cite{Fukugita:2003en,Mohapatra:2005wg,Mohapatra:2006gs,King:2017guk,Feruglio:2019ktm}
and could solve several of its shortcomings \cite{Drewes:2013gca}. Amongst others,\footnote{
	They can e.g.~also be Dark Matter (DM) candidates  \cite{Dodelson:1993je,Shi:1998km}  (cf.~e.g.~\cite{Adhikari:2016bei, Boyarsky:2018tvu} for reviews) or explain oscillation anomalies \cite{Dasgupta:2021ies}.
}
they can explain the masses of ordinary neutrinos via the \emph{seesaw mechanism} \cite{Minkowski:1977sc, GellMann:1980vs, Mohapatra:1979ia, Yanagida:1980xy, Schechter:1980gr, Schechter:1981cv} and
generate the matter-antimatter asymmetry in the observable universe
\cite{Canetti:2012zc}
through \emph{leptogenesis} \cite{Fukugita:1986hr}.
Both, the number $n$ of  flavours $\nu_{R i}$ and their Majorana masses $M_i$ cannot be fixed by established theoretical principles without specifying the UV completion of the SM
and may therefore be treated as free parameters in an agnostic approach.
While  the  original  idea  of  leptogenesis  assumed $M_i$
to be much larger than the electroweak scale, it was realised in the 90s and 2000s that leptogenesis is feasible for $M_i$  below the TeV scale \cite{Pilaftsis:1997jf,Akhmedov:1998qx,Pilaftsis:2003gt,Asaka:2005pn,Pilaftsis:2005rv},
light enough to be produced and discovered at particle accelerators~\cite{Shrock:1980ct,Shrock:1981wq,Gorbunov:2007ak} (see~\cite{Atre:2009rg,Deppisch:2015qwa,Antusch:2016ejd,Cai:2017mow,Chun:2017spz} for reviews).
The exciting possibility to probe the common origin of neutrino mass and the visible matter in the universe has triggered a growing experimental effort to search for heavy neutrinos including searches at the LHC main detectors \cite{Aad:2019kiz,Sirunyan:2018mtv,Sirunyan:2018xiv,Aaij:2020ovh} as well as fixed target expeirments \cite{Abe:2019kgx,NA62:2020mcv}.
Heavy neutrinos have also become an important benchmark model in the physics case studies of proposed future detectors \cite{Beacham:2019nyx,Alimena:2019zri,Agrawal:2021dbo} (including SHiP \cite{Alekhin:2015byh}, MATHUSLA \cite{Curtin:2018mvb}, FASER \cite{Ariga:2018uku}, CODEX-b \cite{Aielli:2019ivi})
and future colliders, such as the FCC-ee \cite{Abada:2019zxq} or CEPC \cite{CEPCStudyGroup:2018ghi}.
From an experimental viewpoint the two most important properties of the heavy neutrinos are their masses $M_i$ and their mixing angles $\theta_{ai}= F_{ai} v /M_i $ with ordinary neutrinos $\nu_{L \alpha}$ (with $v$ the Higgs vev), which determine the strength of the weak interactions that the heavy neutrino mass eigenstates $N_i \sim \nu_{Ri} + \theta_{\alpha i} \nu_{L \alpha}^c$ feel.
The mapping of the viable leptogenesis parameter space and the perspectives to probe it experimentally have been the subject of numerous studies, cf.~e.g.~\cite{Chun:2017spz} for a review.
Exhaustive parameter space scans  so  far  have  almost entirely  been  focused to scenarios that can effectively be described by the model with $n=2$,\footnote{The $\nu$MSM falls into this category because one of the $N_i$ is a DM candidate and has very small couplings, cf.~\cite{Boyarsky:2009ix}.}
the minimal scenario consistent with light neutrino oscillation data. A third $\nu_{R i}$ is not only needed to give mass to the lightest SM neutrino, but
also required in the context of many gauge extensions of the SM to ensure the anomaly freedom of the theory.
In the present work we for the first time identify the range  $U_{\alpha i}^2 = |\theta_{\alpha i}|^2$ for which the extension of the SM by three $\nu_R$ can simultaneously explain the light neutrino masses and the matter-antimatter asymmetry of the universe across the entire experimentally accessible mass range between $50$ MeV and $70$ TeV,\footnote{
For larger masses Eqs.~\eqref{kin_eq} need to be modified to take the equilibration of right-handed electrons into account~\cite{Bodeker:2019ajh}.}
where the lower limit is motivated by the interplay between direct search bounds and cosmological bounds \cite{Chrzaszcz:2019inj} ,\footnote{See e.g.~\cite{Hernandez:2014fha,Vincent:2014rja,Sabti:2020yrt,Boyarsky:2020dzc,Domcke:2020ety,Mastrototaro:2021wzl} for discussions of the cosmological bounds.}
covering the entire experimentally accessible mass range.

\paragraph{\bf Symmetry protected low scale seesaws}
The most general renormalisable extension of the SM by $\nu_R$ only reads
\begin{equation}
	\mathcal L \supset
	\mathrm{i} \overline{\nu_R} \slashed\partial \nu_{R}
	- \frac{1}{2}
	\overline{\nu_R^c}M_M\nu_{R}
	- \overline{\ell_{L}}F \varepsilon \Phi^* \nu_{R}
	+ {\rm h. c.}
	\, , \label{MinimalSeesaw}
\end{equation}
with $F$ a matrix of Yukawa couplings, $M_M$ a Majorana mass matrix, and $\Phi$ and $\ell_L$ the SM Higgs and lepton doublets, respectively.
The model \eqref{MinimalSeesaw} should be regarded as an effective field theory \cite{delAguila:2008ir} with some cutoff scale $\Lambda$, which in principle could be as high as the Planck scale \cite{Bezrukov:2012sa}, but may be considerably lower if it is e.g.~associated with the breaking of additional gauge symmetries.
For instance, in the left-right symmetric model (LRSM) \cite{Pati:1974yy,Mohapatra:1974gc,Senjanovic:1975rk,Senjanovic:1978ev,Minkowski:1977sc,Mohapatra:1979ia} there is a viable corner in parameter space in which the $M_i$ are considerably below the $W_R$ mass \cite{Nemevsek:2018bbt}.
In the present work we remain agnostic regarding the UV completion and simply assume that $\Lambda$ is large enough to justify the use of \eqref{MinimalSeesaw} at all relevant energies and temperatures $T$. We focus on the case with three $\nu_R$ flavours ($n=3$) with all $M_i$ below $10^5$ GeV.\footnote{A discussion of theoretical motivations for a low scale seesaw can e.g.~be found in Sec.~5 of \cite{Agrawal:2021dbo}.}
Though the Lagrangian \eqref{MinimalSeesaw} is identical to the classic high-scale seesaw mechanism, the smallness of the light neutrino masses
in this case
cannot be explained by the ratio $v/M_i$, but instead is typically related to approximate symmetries, which permit to explain the light neutrino masses with $F_{ai}$ of order one for sub-TeV values of the $M_i$.
A popular choice is an approximate $B-\bar{L}$ conservation, with baryon number $B$ and some generalised lepton number $\bar{L}$ under which the $\nu_R$ are charged \cite{Shaposhnikov:2006nn,Kersten:2007vk}.
Popular models that can incorporate this idea e.g.~include the inverse~\cite{Mohapatra:1986aw,Mohapatra:1986bd,Bernabeu:1987gr,GonzalezGarcia:1988rw} and linear~\cite{Akhmedov:1995ip,Akhmedov:1995vm} seesaws, scale invariant models~\cite{Khoze:2013oga} and the Neutrino Minimal Standard Model ($\nu$MSM)~\cite{Shaposhnikov:2006nn} proposed in \cite{Asaka:2005an,Asaka:2005pn}.
A generic feature of these symmetries is the appearance of quasi-degeneracies in the spectrum of $M_i$.
While $B-\bar{L}$ type symmetries typically enforce pair-wise degeneracies amongst the eigenvalues of $M_M$ \cite{Moffat:2017feq}, additional symmetries can imply that all eigenvalues are degenerate  \cite{Pilaftsis:2005rv,Cirigliano:2005ck,Hagedorn:2006ug,Branco:2009by}.
Previous work has shown that the largest $U_{\alpha i}^2$ consistent with leptogenesis for $n=3$ can be achieved in this regime \cite{Abada:2018oly}. Since our goal is to identify the largest and smallest $U_{\alpha i}^2$ for given $M_i$, we focus on the case of three quasi-degenerate $M_i\simeq \bar{M}$ in the following.

In our study of the parameter space we employ the usual Casas-Ibarra parametrization~\cite{Casas:2001sr},
\begin{align}\label{CI}
	F = \frac{i}{v} U_\nu \sqrt{m_\nu^\mathrm{diag}} \mathcal{R} \sqrt{M_M^\mathrm{diag}}\,,
\end{align}
where $U_\nu$ is the PMNS matrix, $m_\nu$ and $M_M$ are the mass matrices of the light and heavy neutrinos respectively.\footnote{
	This parametrization can be modified to include radiative corrections \cite{Pilaftsis:1991ug} by replacing $\sqrt{M_i} \rightarrow M_i/\sqrt{\tilde{M}_i}$~\cite{Lopez-Pavon:2015cga}, as well corrections from unitarity of the mixing matrix~\cite{Donini:2012tt}.
}
The matrix $\mathcal{R}$ satisfies $\mathcal{R} \mathcal{R}^T=1$ and is usually parametrized by three complex Euler angles.
Large imaginary parts of these complex angles correspond to large mixing angles $U^2$ and Yukawa couplings.
In such a parametrisation it is not immediately clear whether the light neutrino masses remain stable under radiative corrections, or how it is related to $B-\bar{L}$ conserving scenarios.
To make this connection more clear, in this work we instead parametrize the matrix $\mathcal{R}$ as:
\begin{align}\label{RParametrisation}
	\mathcal{R} = O_\nu R_C O_N\,,
\end{align}
where $O_\nu=O_\nu^{(13)}O_\nu^{(23)}$ and $O_N=O_N^{(23)} O_N^{(13)}$ are real orthogonal matrices and $R_C$ is the only complex rotation $R_C= R_C^{(12)}$ (where the superscript indicates the plane of rotation).
Intuitively we can understand this parametrization as  rotation
by a complex angle $\omega_c$
around some axis,
which is then rotated to the appropriate basis in $\nu_R$ and $\nu_L$-flavor spaces (we note for completeness that any real $O_N^{(12)}$ and $O_\nu^{(12)}$ can be absorbed into $R_C$, giving us a total of six physical parameters, same as in the complex Euler angle parametrization).
When written this way, a large imaginary part of the angle in $R_C$ leads to a $B-\bar{L}$ conserving structure of the Yukawa couplings, as required by~\cite{Moffat:2017feq}, which is perturbed by the splittings between the eigenvalues of $M_M$.
Explicitly, if we perform a flavor rotation of the Yukawa matrix by $O_N^T$, up to corrections of $\mathcal{O}(|M_i-M_j|)$, and $\mathcal{O}(e^{-|\Im \omega_c|})$, the Yukawa couplings are given by\footnote{
The special case of $n=2$ is realised in the limit $m_\mathrm{lightest}=0$, with $O_N \approx \mathbbm{1}$, where ${\nu_R}_1$ and ${\nu_R}_2$ form a pseudo-Dirac pair and $O_\nu$ is a permutation matrix that depends on the light neutrino mass ordering.}
\begin{align}\label{Fmatrix}
	\tilde{F} &=
	\begin{pmatrix}
		F_{e} & i F_e & 0\\
		F_\mu & i F_\mu & 0\\
		F_\tau & i F_\tau & 0
	\end{pmatrix}, \
	\tilde{M}_M = O_N^T M_M O_N \approx \bar{M}.
\end{align}
If all three neutrinos are quasi-degenerate, the matrix $O_N$ can have significant deviations from the unit matrix, and all three mass eigenstates can have mixings of the same order of magnitude.
In this parametrisation, an approximate mass degeneracy of the three HNLs is a sufficient condition to ensure stability of the light neutrino masses under radiative corrections, irrespective of the choice of the matrix $O_N$.

\paragraph{\bf Low scale leptogenesis}

The simplest thermal leptogenesis scenario requires
$M_i > 10^9$ GeV \cite{Davidson:2002qv}, which can be lowered by taking flavour effects into account \cite{Dev:2017trv}.
Low scale leptogenesis
can be made possible in different ways \cite{Hambye:2001eu}. In the model \eqref{MinimalSeesaw} it is typically achieved by a resonant enhancement
of the contributions from the $N_i$-mode $\textbf{p}$ to the asymmetry generation by a factor
that roughly scales as $\Gamma_{\rm p}/\Delta M$ \cite{Klaric:2021cpi},
where $\Gamma_{\rm p}$ is the thermal width of the $N_i$-mode $\textbf{p}$.\footnote{
	The precise shape of the regulator
	in the limit where $\Delta M^2 = M_i^2 - M_j^2$ vanishes
depends on the scenario under consideration \cite{Dev:2017wwc}.}
In the non-relativistic regime this requires quasi-degeneracies in the spectrum of $M_i$ \cite{Flanz:1994yx,Covi:1996wh},
while in the relativistic regime it is sufficient that $T\gg M_i$ \cite{Drewes:2012ma}.
It is common to distinguish between asymmetries that are generated during the $N_i$'s approach to equilibrium (“freeze-in  mechanism”) and those generated during their decay (“freeze-out mechanism”).
Traditionally the latter is associated with “resonant leptogenesis” \cite{Pilaftsis:2003gt,Pilaftsis:2005rv} while the former is associated with
“baryogenesis from neutrino oscillations” \cite{Akhmedov:1998qx,Asaka:2005pn}, but in general both mechanisms contribute (in particular in the weak washout regime), and oscillations occur during both, freeze-in and freeze-out \cite{Klaric:2021cpi}.

A unified description of all scenarios under consideration here can be achieved with matrix-valued quantum kinetic equations (QKEs)
for the heavy neutrinos and the asymmetries in different SM degrees of freedom,
which can be derived in different ways~\cite{Biondini:2017rpb,Garbrecht:2018mrp}. In the present work we use the QKEs
presented in \cite{Klaric:2021cpi},
\begin{subequations}
	\begin{align}
		i \frac{d n_{\Delta_\alpha}}{dt}
		&= -2 i \frac{\mu_{\alpha}}{T} \int \frac{d^{3} k}{(2 \pi)^{3}} \operatorname{Tr}\left[\Gamma_{\alpha}\right] f_{N}\left(1-f_{N}\right) \nonumber\\
		&\quad +i \int \frac{d^{3} k}{(2 \pi)^{3}} \operatorname{Tr}\left[\tilde{\Gamma}_{\alpha}\left(\bar{\rho}_{N}-\rho_{N}\right)\right],
		\label{kin_eq_a}
		\\
		i \, \frac{d\rho_{N}}{dt}
		&= \left[H_{N}, \rho_{N}\right]-\frac{i}{2}\left\{\Gamma, \rho_{N}-\rho_{N}^{eq} \right\} \nonumber\\
		&\quad-\frac{i}{2} \sum_{\alpha} \tilde{\Gamma}_{\alpha}\left[2 \frac{\mu_{\alpha}}{T} f_{N}\left(1-f_{N}\right)\right] ,
		\label{kin_eq_b}
		\\
		i \, \frac{d \bar{\rho}_{N}}{d t}
		&= -\left[H_{N}, \bar{\rho}_{N}\right]-\frac{i}{2}\left\{\Gamma, \bar{\rho}_{N}-\rho_{N}^{eq} \right\} \nonumber\\
		&\quad+\frac{i}{2} \sum_{\alpha} \tilde{\Gamma}_{\alpha}\left[2 \frac{\mu_{\alpha}}{T} f_{N}\left(1-f_{N}\right)\right].
		\label{kin_eq_c}
\end{align}\label{kin_eq}\end{subequations}
Here $\rho_{N}$  and $\bar{\rho}_{N}$ are the momentum averaged density matrices for the two $N_i$ helicities, $f_N$ is the Fermi-Dirac distribution, and $\mu_{\alpha}$ are flavoured lepton chemical potentials that are related to their comoving number densities $n_{\Delta_\alpha}$ by a susceptibility matrix \cite{Buchmuller:2005eh}. 
$H_N$ is an effective Hamiltonian for the $N_i$, and  $\Gamma$, $\Gamma_\alpha$ and $\tilde{\Gamma}_\alpha$ are various thermal interaction rates taken from \cite{Ghiglieri:2017gjz}. 

\paragraph{\bf Results}

We solve the set of equations \eqref{kin_eq} for two types of initial conditions for the $N_i$ : $i)$ vanishing abundances or $ii)$ thermal initial abundances. 
The former apply in scenarios where the reheating temperature is considerably lower than $\Lambda$, such as the $\nu$MSM, the latter apply in scenarios where the $\nu_R$ have additional interactions at high energies and the reheating temperature is larger than $\Lambda$.
In both cases we assume that all SM chemical potentials vanish initially, cf.~\cite{Asaka:2017rdj,Domcke:2020quw} for a recent discussion of this point.

The model \eqref{MinimalSeesaw} contains $7n-3$ free parameters, 
and the equations \eqref{kin_eq} contain a large number of times scales that can depend on all of them,
making a complete exploration of the parameter space numerically challenging.
The goal of the present work is a first identification of the range of mixings $U_{\alpha}^2=\sum_i U_{\alpha i}^2$ 
and $U^2=\sum_\alpha U_\alpha^2$
for given $\bar{M}$ for which a matter-antimatter asymmetry that exceeds the current observational value of the baryon-to-photon ratio \cite{Akrami:2018vks} can be generated while simultaneously explaining the light neutrino oscillation data without fine-tuning.  
For this purpose we focus on the case of normal light neutrino mass ordering and separately consider the two choices
$m_{\rm lightest}=0\mbox{~eV}$ and $m_{\rm lightest} = 0.1$ eV.
We fix all measured neutrino oscillation parameter to the best fit values presented in \cite{Esteban:2020cvm} and used uniform priors for the Dirac and Majorana phases in $U_\nu$ as well as the real parts of the rotation angles in $\mathcal{R}$. For the imaginary part of the rotation angle in $R_C$ we used a uniform prior between $-10$ and $10$. 
In each step of the scan we fixed $\bar{M}$ to a particular value and randomised the relative deviations of the eigenvalues in $M_M$ with uniform $\log_{\rm 10}$ priors between $-10$ and $-1$.

We impose three criteria on all parameter points to ensure the theoretical consistency of the computations. 
\emph{Perturbative unitarity}: For the $N_i$ to be well defined as quasiparticles, we require their decay width to be smaller than half their mass.
\emph{Seesaw expansion}: We only consider points with $U^2<0.1$ to ensure that the parameterisation \eqref{CI} is approximately valid.
\emph{No fine tuning}. We require small radiative corrections for the light neutrinos masses which suggests the presence of an underlying $B-\bar{L}$ symmetry. As discussed after \eqref{RParametrisation}, this condition is automatically satisfied when all $M_i$ are quasi-degenerate.

\begin{figure}[ht!]
	\centering
	\begin{subfigure}[b]{\columnwidth}
		\includegraphics[width = \linewidth]{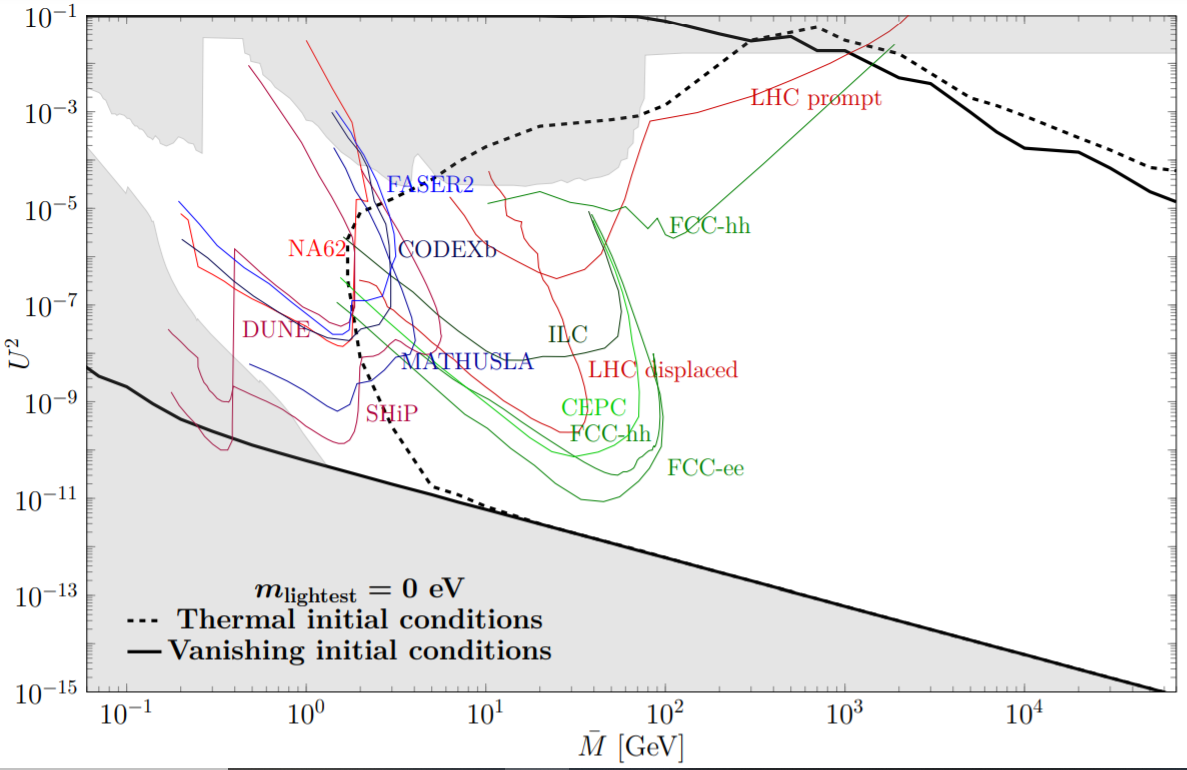}
	\end{subfigure}
	\\
	\begin{subfigure}[b]{\columnwidth}
		\includegraphics[width = \linewidth]{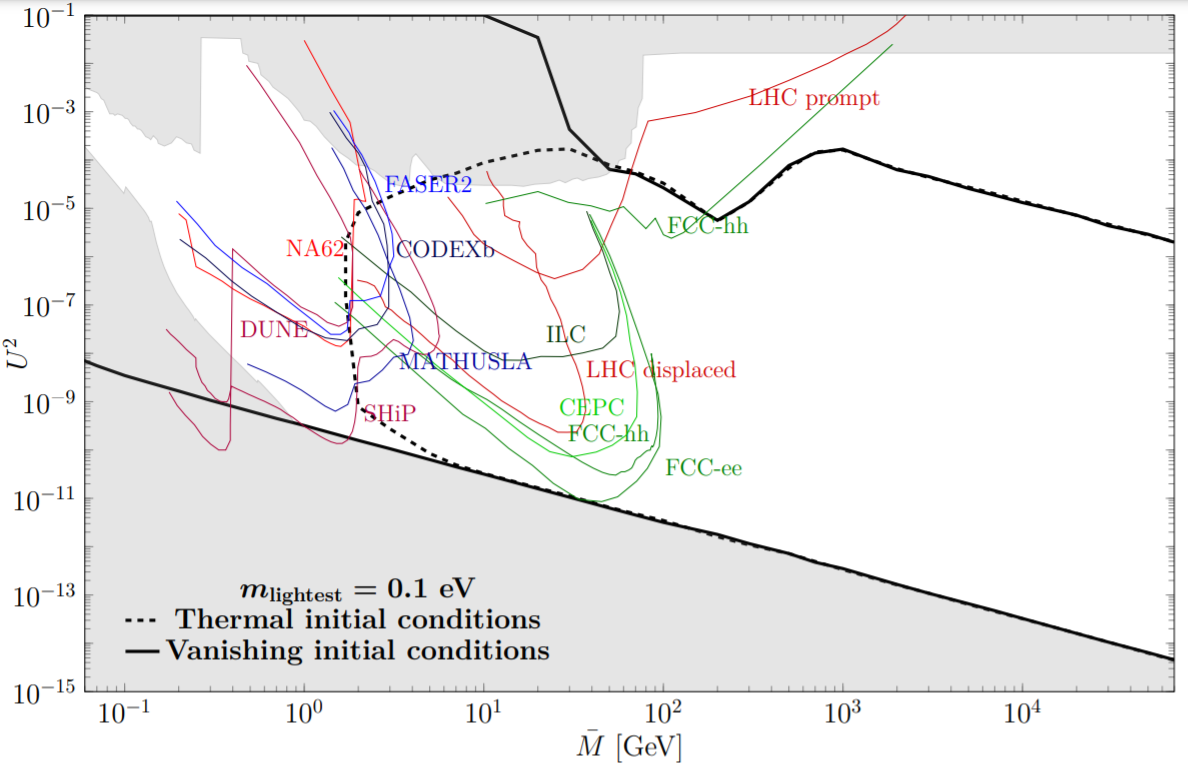}
	\end{subfigure}
	\caption{Allowed parameter space for leptogenesis with 3 heavy neutrinos for vanishing (inside solid black line) and thermal (inside dashed black line) initial conditions and $m_{\rm lightest}=0$ eV (upper panel) or $m_{\rm lightest}=0.1$ eV (lower panel). 
		The gray area indicates the experimentally excluded region identified in the global scan \cite{Chrzaszcz:2019inj}, complemented by the updated BBN bounds from~\cite{Sabti:2020yrt,Boyarsky:2020dzc}.
		The coloured lines indicate the estimated sentitivities of the LHC main detectors (taken from \cite{Izaguirre:2015pga,Drewes:2019fou,Pascoli:2018heg}) and 
		NA62 \cite{Drewes:2018gkc} along with that of selected planned or proposed experiments (DUNE \cite{Ballett:2019bgd},
			FASER2 \cite{Ariga:2018uku},
			SHiP \cite{SHiP:2018xqw,Gorbunov:2020rjx}
			MATHUSLA \cite{Curtin:2018mvb},
		Codex-b \cite{Aielli:2019ivi}) as well as
		future lepton colliders \cite{Antusch:2017pkq} or proton colliders \cite{Antusch:2016ejd}. 
	}
	\label{results}
\end{figure}

The results of this scan are displayed in Fig.~\ref{results}. As it is the most important quantity from a phenomenological viewpoint, we show the results in the $\bar{M}-U^2$ plane in this Letter. We will present other representations of the $18$-dimensional parameter space and a detailed discussion in a longer follow-up work. 
In Fig.~\ref{results} we compare this to the reach of selected experiments in the $\bar{M}-U_\mu^2$ plane. 
This is justified because we find that the maximal allowed $U_\mu^2$ tends to be very close to the maximal $U^2$ for given $\bar{M}$. In contrast to that, the maximal $U_e^2$ in the case $m_{\rm lightest}=0$ is about an order of magnitude smaller, which can be understood from Fig.~11 in \cite{Chrzaszcz:2019inj} and is consistent with the known fact that the range of $U_e^2/U^2$ allowed by neutrino oscillation data is small for $n=2$ \cite{Hernandez:2016kel,Drewes:2016jae,Caputo:2017pit,Drewes:2018gkc}.

\paragraph{\bf Discussion and conclusion}
Our main findings are as follows:
\begin{itemize}
	\item[1)] For both types of initial conditions we find that leptogenesis with $n=3$ is possible with $U^2$ that are orders of magnitude larger than in the case $n=2$ \cite{Klaric:2020lov,Klaric:2021cpi}.
\item[2)] For thermal initial conditions leptogenesis is feasible for masses as low as $1.7$ GeV and $U^2$ that are accessible to current experiments. For vanishing initial conditions the lower bound on $\bar{M}$ from leptogenesis is weaker than that from experiments and BBN.
	\item[3)] The maximal $U^2$ is larger for $m_{\rm lightest}=0$ eV than for $m_{\rm lightest}=0.1$ eV.
\item[4)] For $m_\text{lightest} = 0$ and $\bar{M}>10^2$ GeV, we find that leptogenesis from thermal initial conditions can
lead to larger mixing angles than leptogenesis with 
vanishing initial $N_i$ occupation numbers.
\end{itemize}
One of the most surprising outcomes of the parameter scan is the observation $1)$ that the allowed mixing angles in the scenario with three heavy neutrinos exceed the leptogenesis bounds from the scenario with $n=2$ \cite{Klaric:2021cpi} by several orders of magnitude.
A similar observation has been made in~\cite{Abada:2018oly}, where only the freeze-in and $\bar{M}$ below $50$ GeV were considered. 
Amongst the various differences between $n=2$ and $n=3$ discussed in \cite{Abada:2018oly}, two are most relevant here.
$i)$ Lepton asymmetries can be preserved from large washout by a \emph{flavor hierarchical washout}, since the ratios $U_{\alpha i}^2/U_i^2$ with $n=3$ are much less constrained by neutrino oscillation data than for $n=2$ \cite{Chrzaszcz:2019inj}.
$ii)$ Thermal effects can cause a level-crossing between the $N_i$ dispersion relations (similar to the well-known MSW effect) that resonantly enhances the asymmetry production, which cannot be realised in the $B-\bar{L}$ limit for $n=2$.
For the freeze-out (where the $N_i$ are non-relativistic) these two effects appear to play a much smaller role since we find a large population of points that do not satisfy either of these two criteria.
We instead find that a crucial element in preventing washout is that one direction in the $\nu_{R i}$ flavor space can remain weakly coupled   and can have a much more significant deviation from equilibrium. This is in contrast to the case with $n=2$, where both reach equilibrium soon after they become non-relativistic because they form a pseudo-Dirac pair.
The deviation from equilibrium during decays is typically of the order $\delta n_i \approx \dot{n}^\mathrm{eq}/\Gamma$, where $\Gamma \approx \Gamma_1 \approx \Gamma_2$ is the inverse lifetime of the two neutrinos $\nu_{R1}$ and $\nu_{R2}$ that form the pseudo-Dirac pair with $M_2\simeq M_3$.
If we include a third neutrino $\nu_{R3}$, its lifetime is not necessarily determined by the mixing angle $U^2$, it can have a much bigger deviation from equilibrium.
If $M_3$ is very different from $M_2$ and $M_1$, the $B-\bar{L}$ symmetry dictates that $\nu_{R1}$ and $\nu_{R2}$ form a pseudo-Dirac pair of mass eigenstates $N_i$ (first two columns in \eqref{Fmatrix}) with mixings of order $U^2$, while the third mass eigenstate $N_3$ remains feebly coupled (third column in \eqref{Fmatrix}).
However, in the triple mass-degenerate scenario, $\nu_{R3}$ can mix with the pseudo-Dirac pair through the mass term.
This explains not only point $1)$, but also point $3)$ because smaller $m_{\rm lightest}$ allow for smaller couplings of $\nu_{R 3}$. 
Regarding $2)$, leptogenesis with thermal initial conditions is possible for $\bar{M}\ll v$ because the enhancement of the asymmetry due to resonant and flavour effects can be sufficient to overcome the suppression by $(\bar{M}/T)^2$ of the deviation from equilibrium~\cite{Hambye:2016sby,Granelli:2020ysj,Klaric:2020lov,Klaric:2021cpi}.\footnote{
	The late decay of the $N_i$ in this scenario could potentially generate a lepton asymmetry that greatly exceeds the baryon asymmetry, which can have interesting phenomenological consequences, including enhanced singlet fermion DM production \cite{Shi:1998km} and affect the nature of the QCD transition \cite{Schwarz:2009ii} and primordial black hole production~\cite{Bodeker:2020stj}.
}
Finally, point $4)$ is a result of the well-known fact that the asymmetries generated during freeze-in and freeze-out have opposite signs \cite{Buchmuller:2004nz} (cf.~\cite{Garbrecht:2019zaa} for a recent discussion) and partially cancel each other in the case of vanishing initial conditions. 

The much larger range of masses and mixings for which leptogenesis is feasible for $n=3$ compared to $n=2$ do not only imply considerably better chances for existing experiments to discover the $N_i$, but also imply that a much larger number of them may be observed. 
The price at which this comes is the larger number of model parameters, which makes the model with $n=3$ less predictive than with $n=2$, where in principle all model parameters can be constrained experimentally \cite{Hernandez:2016kel,Drewes:2016jae}. 
In spite of this, with such a large number of events, one can perform several consistency checks of the hypotheses that the model \eqref{MinimalSeesaw} can simultaneously generate the light neutrino masses and the matter-antimatter asymmetry in the universe. 
For instance, if $U^2$ happens to lie near the current experimental limit, we estimate (using the results of \cite{Drewes:2019fou}) that the HL-LHC could observe thousands of displaced vertex events.
This would permit a percent level determination of the fractions  $U_\alpha^2/U^2$ (cf.~appendix B of \cite{Antusch:2017pkq}). Moreover, 
the amount of $B-\bar{L}$ breaking can be studied by
measuring the total ratio between lepton number violating and conserving decays \cite{Anamiati:2016uxp,Anamiati:2017rxw,Das:2017hmg,Dib:2017iva,Abada:2019bac,Drewes:2019byd} (cf.~also~\cite{Boyanovsky:2014una,Cvetic:2015ura}),
the dependence of this quantity on $\Gamma$ \cite{Antusch:2017ebe,Cvetic:2018elt,Tastet:2019nqj},
the $U_\alpha^2/U^2$ \cite{Dib:2016wge},
or the $N_i$ momentum distribution \cite{Arbelaez:2017zqq,Balantekin:2018ukw,Blondel:2021mss}.
Finally,  parameters that are not directly accessible ($\mathcal{R}$, Majorana phases in $U_\nu$)
may be accessed by measuring the CP-violation
in $N_i$ decays \cite{Cvetic:2015naa},  the $U_\alpha^2/U^2$ \cite{Drewes:2016jae,Caputo:2016ojx} or neutrinoless double $\beta$-decay \cite{Drewes:2016lqo,Hernandez:2016kel,Asaka:2016zib,Dekens:2020ttz,Harz:2021psp}.

In summary, we for the first time studied the range of mixing angles $U_\alpha^2$ for which right-handed neutrinos with Majorana masses below $100$ TeV in the model \eqref{MinimalSeesaw} can generate the light neutrino masses and the baryon asymmetry of the universe. We considered both, the cases of vanishing initial conditions that one would find in minimal models like the $\nu$MSM and thermal initial conditions that one would expect in models with an extended gauge sector. We find that this range is much bigger than in the case of two right-handed neutrinos, which opens up the possibility to see thousands of events in existing experiments. When combined with data from neutrino oscillation experiments and neutrinoless double $\beta$-decay, this permits various consistency checks to test the hypothesis that right-handed neutrinos are common the origin of neutrino masses and baryonic matter in the universe

\begin{acknowledgments}
	\paragraph{Acknowledgments}
	We would like to thank Claudia Hagedorn for helpful discussions on flavour symmetries
	as well as Oliver Fischer for comments on collider constraints and Oleg Ruchayskiy for comments  on BBN. We also thank 
	Pasquale Di Bari,
	Mikhail Shaposhnikov and Inar Timiryasov  for proof-reading this manuscript.
	JK acknowledges the support of the ERC-AdG-2015 grant 694896.
\end{acknowledgments}

\raggedright\bibliography{inspire_bibiliography}

\end{document}